\RequirePackage{booktabs}

\documentclass[sn-mathphys-num]{sn-jnl}% Math and Physical Sciences Numbered Reference Style 
\usepackage{graphicx}%
\usepackage{multirow}%
\usepackage{amsmath,amssymb,amsfonts}%
\usepackage{amsthm}%
\usepackage{mathrsfs}%
\usepackage[title]{appendix}%
\usepackage{xcolor}%
\usepackage{textcomp}%
\usepackage{manyfoot}%
\usepackage{booktabs}%
\usepackage{algorithm}%
\usepackage{algorithmicx}%
\usepackage{algpseudocode}%
\usepackage{listings}%

\usepackage{lmodern}
\usepackage{tabularx}
\usepackage{caption}
\usepackage{lscape}
\usepackage{lineno}

\usepackage{float}

\begin{document}

\title[Article Title]{Milgram's experiment in the knowledge space: Individual navigation strategies}

\author*[1,2]{\fnm{Manran} \sur{Zhu}}\email{manran.zhu@uni-corvinus.hu}

\author[1]{\fnm{János} \sur{Kertész}}\email{KerteszJ@ceu.edu}

\affil*[1]{\orgdiv{Department of Network and Data Science}, \orgname{Central European University}, \orgaddress{\street{Quellenstraße 51}, \city{Vienna}, \postcode{1100}, \state{Vienna}, \country{Austria}}}

\affil[2]{\orgdiv{Center for Collective Learning, CIAS}, \orgname{Corvinus University of Budapest}, \orgaddress{\street{Kozraktár u. 4-6}, \city{Budapest}, \postcode{1093}, \state{Budapest}, \country{Hungary}}}

\abstract{Data deluge characteristic for our times has led to information overload, posing a significant challenge to effectively finding our way through the digital landscape. Addressing this issue requires an in-depth understanding of how we navigate through the abundance of information. Previous research has discovered multiple patterns in how individuals navigate in the geographic, social, and information spaces, yet individual differences in strategies for navigation in the knowledge space has remained largely unexplored. To bridge the gap, we conducted an online experiment where participants played a navigation game on Wikipedia and completed questionnaires about their personal information. Utilizing the hierarchical structure of the English Wikipedia and a graph embedding trained on it, we identified two navigation strategies and found that there are significant individual differences in the choices of them. Older, white and female participants tend to adopt a proximity-driven strategy, while younger participants prefer a hub-driven strategy. Our study connects social navigation to knowledge navigation: individuals’ differing tendencies to use geographical and occupational information about the target person to navigate in the social space can be understood as different choices between the hub-driven and proximity-driven strategies in the knowledge space. }




\maketitle

\section{Introduction}\label{Introduction}
Navigating from one place to another is a crucial ability for animals, enabling them to locate essential resources such as food, mates, and habitats~\cite{epstein2017cognitive, stachenfeld2017hippocampus}. Seeking resources occurs not only in the physical space but also in more abstract spaces, such as when we look for the right person for assistance in the social space~\cite{tavares2015map, schafer2018navigating}, or when searching for an answer to a question online in the knowledge space~\cite{pirolli1999information}. With the accumulation of massive information online in the past decades, information overload has become a significant challenge for our generation, making efficient way-finding in the information space crucial~\cite{bush1945we}. To tackle this challenge, we first need to understand how we navigate the information space. 

Studies of navigation behavior originated in the physical domain, where the cognitive map theory~\cite{tolman1948cognitive, whittington2022build} was developed to explain how humans and animals mentally represent spatial environments and determine routes within them. In recent years, this theory has been extended beyond physical space to encompass navigation in abstract domains, including social and informational spaces~\cite{epstein2017cognitive}. Notably, research has shown that “concept cells” in the hippocampus and entorhinal cortex—originally thought to encode spatial locations—also represent abstract concepts and social relationships, suggesting shared neural substrates across domains~\cite{son2021cognitive}. Social navigation (e.g., inferring social hierarchies or choosing allies) and information navigation (e.g., exploring knowledge networks or digital content) rely on similar cognitive processes, such as mapping, orientation, and decision-making under uncertainty. Understanding the connections between these forms of navigation can thus reveal fundamental principles of how the brain organizes, accesses, and utilizes complex, high-dimensional information. Such insights are crucial for designing better information systems, enhancing learning, and modeling human behavior in increasingly digital environments. 

Previous research have found that our efficient navigation ability in social space is linked to the structure of the social network~\cite{watts2002identity, kleinberg2001small}. The way we are connected socially is highly structured: we all possess different identities~\cite{white1992identity} and belong to groups characterized by specific social attributes~\cite{simmel1902number, nadel2013theory}. These group structures naturally form hierarchies, akin to the departmental organization in universities or companies, where individuals belong to groups, which in turn belong to larger groups. Watts et al. \cite{watts2002identity} and Kleinberg~\cite{kleinberg2001small} proved that networks equipped with a hierarchical structure are navigable: utilizing a greedy decentralized search algorithm where one always chooses the next step to be the one that's closest to the target, one can navigate to any target person in a small number of steps. This theory was later empirically confirmed by Adamic et al.~\cite{adamic2005search} who demonstrated that given the email logs a greedy decentralized search could effectively leverage the organizational hierarchy to find short paths to the target. In fact, different hierarchies can be utilized: in the context of social navigation, individuals typically rely on either geographical or occupational hierarchies to facilitate their navigation~\cite{dodds2003experimental, killworth1978reversal}. 

Does social navigation theory applies to navigation in the knowledge space? Targeted navigation in the information space similar to the Milgram's experiment has been implemented and studied in the game setting on the Wikipedia~\cite{Wikipedia}, noted for its wide topic range and significant user interaction. In the popular Wikipedia navigation games such as Wikispeedia~\cite{Wikispeedia} and the Wiki Game~\cite{Wikigame}, players are challenged to move from one Wikipedia article (source page) to another (target page) along a chain of hyperlinks of the visited Wikipedia articles. Researchers have uncovered intriguing patterns in the navigation game. Analysing 30,000 instances of players' navigation paths, West et. al.~\cite{west2012human} discovered that the players' navigation typically consists of two phases: the zoom-out phase in the early game when players tend to visit high degree node (hub); and the home-in phase when players constantly decrease the conceptual distance to the target. Looking at the navigation decision players made at each step, studies have found that players' navigation decisions are biased and stochastic: their current navigation decisions are biased by their previous decisions on the topical level~\cite{singer2014detecting}, and they sometimes randomly select their next moves, particularly in the early stages of the game~\cite{helic2013models}, which presents a trade-off of exploration and exploitation in searching behavior~\cite{hills2015exploration, brin1998anatomy, coulom2006efficient}.

While previous research has shed light on how we navigate in general, a comprehensive understanding of how we navigate differently remains elusive. Studies on the cognitive social structure~\cite{krackhardt1987cognitive, brands2013cognitive, smith2020social} have shown that we humans are very good at “filling in the blanks”: from our observation of social interactions, we tend to infer the interactions that are not directly observed and form an abstract representation (schema) of the social network that is highly structural, with categories and hierarchies~\cite{freeman1992filling}. These schemas are biased across individuals~\cite{killworth1976informant, bernard1977informant} and can lead to different social navigation behaviors~\cite{russell1982index}, and even social status differences: researchers found that people with a more accurate cognition of the advice network in the firm are rated as more powerful by their peers~\cite{krackhardt1990assessing}. Aside from the schema differences, our status, power, and emotions all affect our navigation behavior: faced with a job threat, people experiencing low status and negative emotions tend to activate smaller and denser social networks to search for a job, while people experiencing the opposite tend to activate larger and sparser networks~\cite{smith2012status}.

Moving from social navigation to navigation in the information space, understanding individual differences faces several challenges: although data for information foraging on the World Wide Web is abundant, raw web request logs with user's information such as their IP address are considered very sensitive and therefore usually not available~\cite{arora2022wikipedia}. Online games such as the Wikispeedia and Wikigame produce massive amount of user navigation trails, but the absence of participants' demographic information hampers the exploration of how individual traits impact navigation patterns. What's more, despite extensive research on social navigation and knowledge navigation in their respective domain, an integrated understanding of the navigation strategies adopted in both processes is still lacking. To overcome these limitations, we conducted an online experiment where we hired 802 participants online from the United States to play nine rounds of the Wikipedia navigation games and then complete a survey (for details, see section \ref{subsec:experiment}) that included questions about their demographic information and other factors potentially relevant to their navigation behavior. Building on the insights gained from previous studies on social navigation, we tailored our experiment to focus on navigation between social persons within the information space: the source and target pages for the navigation games in our study were selected to be well-known individuals from various professions, genders, and historical periods (see Fig. \ref{fig:hc_paths} for a list of the source and target pages of each game). 

Following previous observations that we utilize both the distance and hierarchy structure of the knowledge space to navigate, we trained a graph embedding on the English Wikipedia network to quantify the pairwise distance among the Wikipedia pages and calculated a hierarchical score for each Wikipedia page to measure its position in the knowledge hierarchy. We found that the split between hub-driven and proximity-driven tendencies is not only present within a single navigation game characterized by the zoom-out and home-in phases~\cite{west2012human}, but also at the individual level. This individual variance is statistically significant and cannot be overlooked. Demographic factors influence not only navigation performance, as demonstrated in our previous work~\cite{zhu2023individual}, but also navigation strategies. Our study further connects social navigation to knowledge network navigation: individuals’ differing tendencies to use geographical and occupational information about the target person can be understood as different choices between hub-driven and proximity-driven strategies.

\section{Results}\label{Results}

\subsection{Navigation paths in the knowledge space}
Previous studies have shown that we utilize the semantic and hierarchical structure of the knowledge space to navigate~\cite{helic2013models}. Here we provide a more detailed picture of the participants' navigation patterns with regards to the two structures. To quantify the distance between two Wikipedia articles, we trained a 64-dimensional embedding using the DeepWalk algorithm~\cite{perozzi2014deepwalk}, which maps each Wikipedia article $a_i$ to a vector $\vec{v}_i$ in the embedding space where more related articles are placed closer together (see section \ref{subsec:embedding} for more details). This method allows us to measure how ``close" two articles $a_i$ and $a_j$ are by the cosine similarity of their respective vectors $\vec{v}_i$ and $\vec{v}_j$ in the embedding space (Eq. \ref{eq:c}). In particular, the closeness score $c(a_i)$ of the article $a_i$ relative to the target Wikipedia page $a_{target}$ of the game is calculated as $c(a_i) = c(a_i, a_{target})$. 

\begin{equation} 
\label{eq:c}
    c(a_i, a_j) = 1 + cos(\vec{v}_i, \vec{v}_j)
\end{equation}

Aside from distance, hierarchy can also be extracted from a network. Muchnik et al.~\cite{muchnik2007self} developed a local hierarchy measure, $h$, which represents an article’s position within Wikipedia’s knowledge hierarchy. Calculated from the in-degree $k_{in}(i)$ and out-degree $k_{out}(i)$ of the article $a_i$ on the Wikipedia network (Eq.~\ref{eq:h}), this local hierarchy measure was validated as an invariant across different language editions of Wikipedia and performs comparably to more global hierarchy measures, such as hierarchical intermediacy and the attraction-basin hierarchy measure~\cite{muchnik2007self}.

\begin{equation} 
\label{eq:h}
    h(a_i) = \frac{k_{in}^{3/2}(i)+k_{out}^{3/2}(i)}{k_{in}(i) + k_{out}(i)} 
\end{equation}

\begin{figure}[H]
    \centering
    \includegraphics[width=\textwidth]{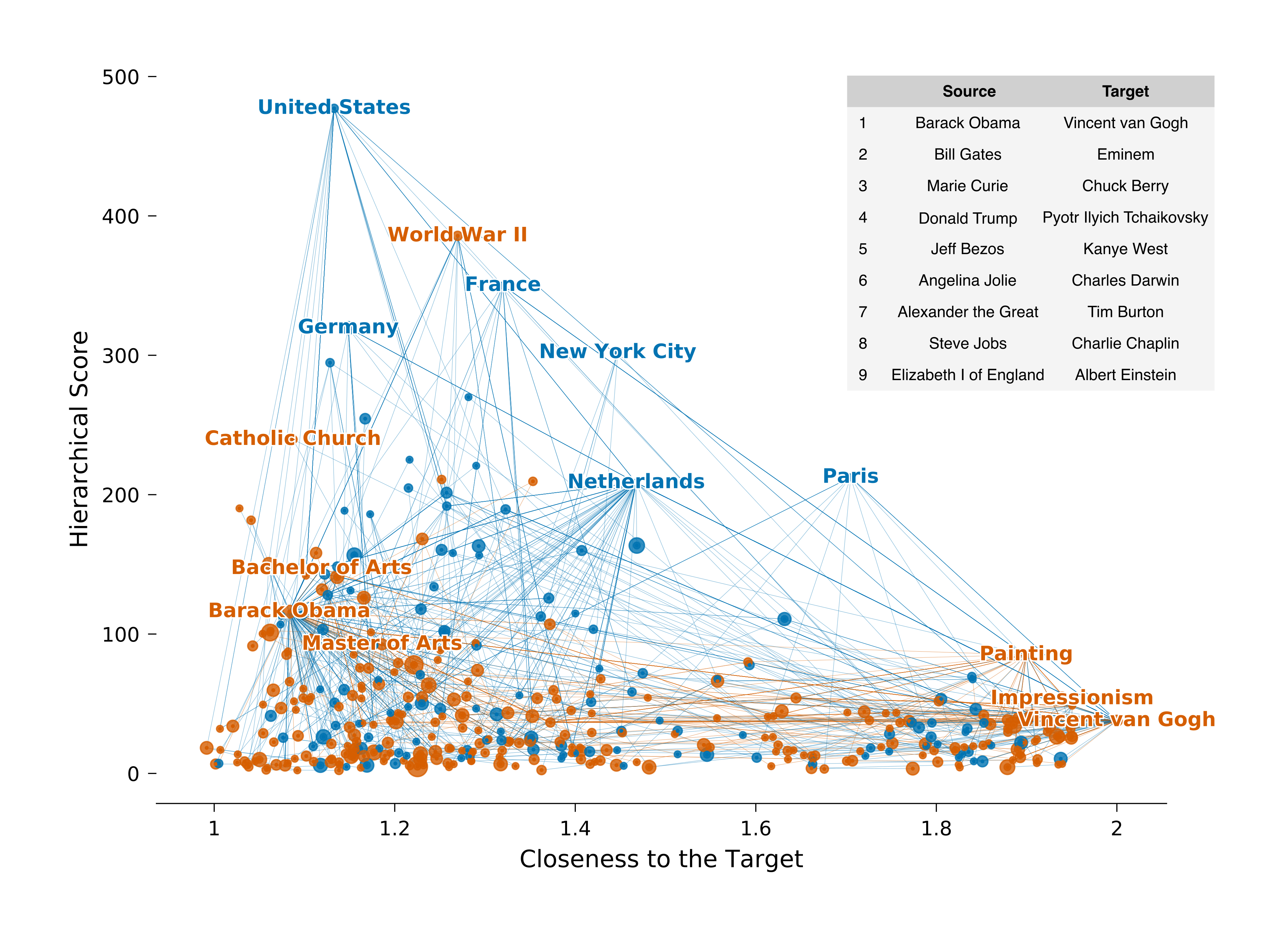}
    \caption[The figure shows all the successful navigation paths and the visited Wikipedia articles for the game with ``Barack Obama" as the source and ``Vincent van Gogh" as the target. The horizontal axis shows the articles' closeness to the target ($c$ in Eq. \ref{eq:c}), and the vertical axis shows the hierarchical score ($h$ in Eq. \ref{eq:h}). Geography-related articles and paths (see definition in section \ref{subsec:interplay}) are marked in blue; others in red. Dot size is proportional to the logarithm of the total number of visits to each article. The top-right table lists source and target articles for all nine games.]{Visualization of the successful navigation paths. }
    \label{fig:hc_paths}
\end{figure}

Fig. \ref{fig:hc_paths} illustrates all successful navigation paths for the game with the source page ``Barack Obama" and the target page ``Vincent van Gogh" in terms of the hierarchical score $h$ and closeness score $c$ of the articles in the paths. As shown, some navigation paths are more ``hub-driven" by ascending higher in the hierarchy to reach hubs before descending towards the target, while others are more ``proximity-driven", by maintaining a lower profile, aiming to minimize their distance to the target. This closely parallels the two navigation phases discovered previously~\cite{west2012human}: a zoom-out phase in the early period of the game, characterized by the players' tendency to visit high-degree nodes (hubs), and a home-in phase, when players consistently decrease the conceptual distance to the target. In the next section, we will show quantitatively that this difference in tendency present during the two phases in each game is also present at the individual level. 

%fig:hc_paths

\subsection{Hub-driven and proximity-driven strategies}
\label{subsec:HC}

The distinction between the hub-driven and proximity-driven navigation paths can be elucidated better through an analogy with transportation networks. In road networks, where shortcuts between destinations are limited, travelers often rely on a proximity-driven strategy, targeting at locations that are close to their final destination due to the necessity of traversing adjacent locations. Conversely, in networks rich with shortcuts, such as airline networks, a hub-driven strategy becomes more viable. Hubs, despite not necessarily being close to the final destination, offer extensive connections across numerous locations.

To quantify to what extent a navigation path is proximity-driven or hub-driven, we calculated a hub-driven score $H$ and proximity-driven score $C$ for each navigation path as the average hierarchical score $h$ or closeness to the target score $c$ of each article in the path. To be more specific, for a navigation path $j$ consisting of articles $A_j = \{a_k\}$, the hub-driven score $H(j)$ and proximity-driven score $C(j)$ are defined as:

\begin{align}
\label{eq:CH}
H_{j} &= \frac{1}{N_{j}} \sum_{a_k \in A_j} h(a_k) \\
C_{j} &= \frac{1}{N_{j}} \sum_{a_k \in A_j} c(a_k, target)
\end{align}

Where $N_j = |A_j|$ is the number of the articles in the path, $h(a_k)$ the hierarchical score (Eq. \ref{eq:h}) of article $a_k$, and $c(a_k, target)$ the closeness to the target page (Eq. \ref{eq:c}). To make the scores comparable across all nine games, we linearly scaled $H$ and $C$ to the range $[0, 1]$ using the min-max scaler. 

\begin{figure}[H]
    \centering
    \includegraphics[width=\linewidth]{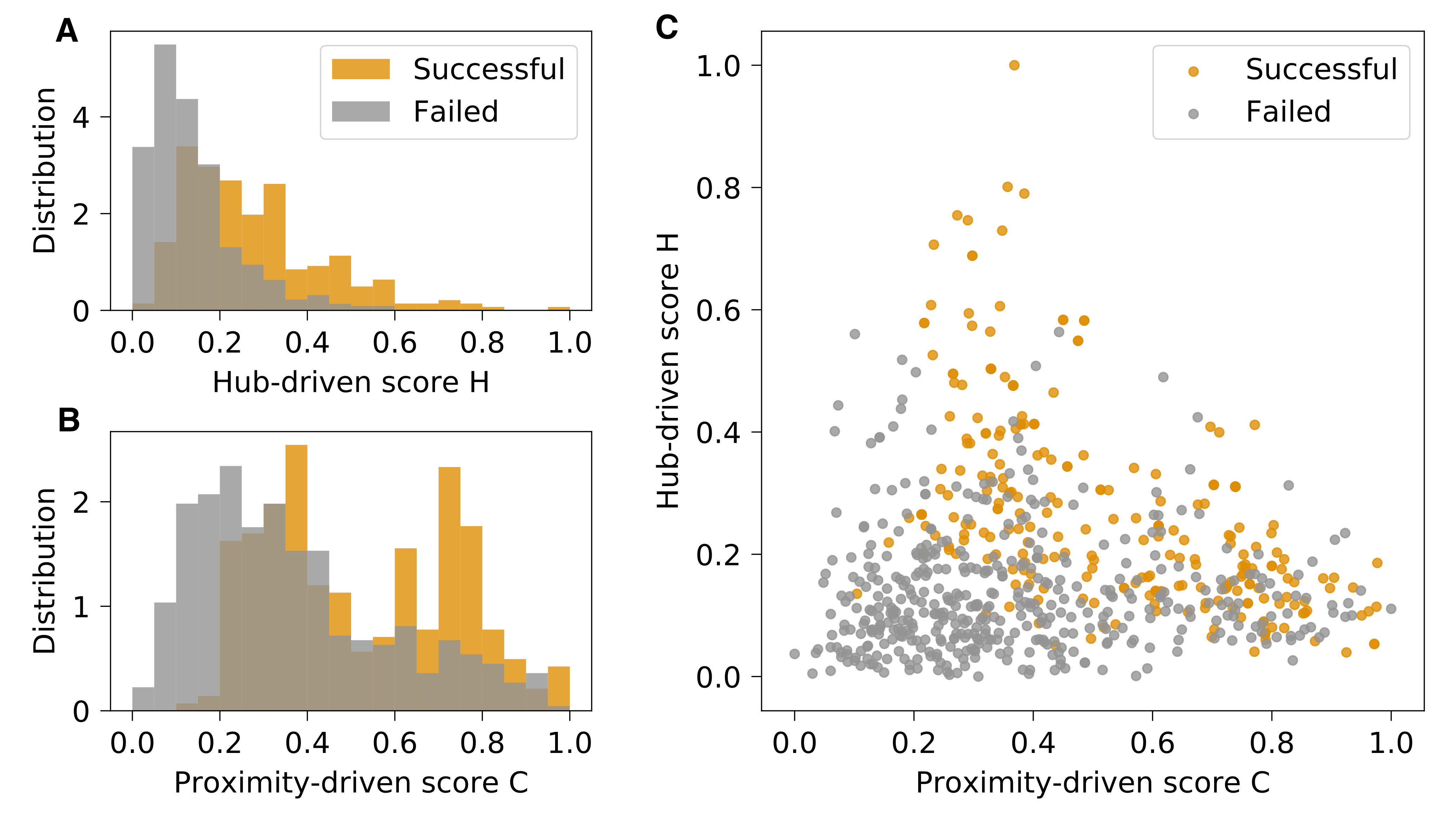}
    \caption[Figure \textbf{A} and \textbf{B} show the distribution of the hub-driven score $H$ and proximity-driven score $C$ for successful and failed navigation paths, respectively, in the game with ``Barack Obama" as the source and ``Vincent van Gogh" as the target. Figure \textbf{C} shows the relationship between $H$ and $C$, with each dot representing a navigation path. ]{Distribution of the hub-driven and proximity-driven scores. }
    \label{fig:hc_distribution}
\end{figure}

Fig. \ref{fig:hc_distribution} visualizes the distribution and the relationship of the hub-driven and proximity-driven scores for each navigation path, with successful paths marked in orange and failed paths in gray. Among successful paths, we observe a negative correlation between the two scores across all nine games (M = -0.55, SD = 0.10), suggesting a trade-off between the two navigation strategies: while both approaches can lead to success, prioritizing one typically requires compromising the other. For failed paths, we identify two distinct patterns. A substantial proportion of failures correspond to paths with both low $H$ and low $C$ scores, indicating that players did not find useful clues during navigation. Other failed paths exhibit score distributions similar to those of successful paths, suggesting that these attempts were close to completion but ultimately unsuccessful due to constraints on time or number of steps.

Is hub-driven approach more effective than proximity-driven approach? To address this question, we implemented linear regression models to predict the performance of the players measured by the time (in seconds) and steps saved in the Speed-race and Least-clicks respectively. Here, we focus solely on successful navigation paths, positing that minimizing time and steps reflects superior performance. To control for other factors affecting players' performance, we included the following covariates: a categorical variable \( Game \), indicating which of the nine games was played; \( Round \), an integer (1–9) representing the game round; two variables for participants' self-reported prior knowledge of the source and target articles; and two performance measures, \( Steps \) and \( Seconds \), representing the total number of steps taken and the duration of the game, respectively.

Table \ref{tab:performance} shows that the effectiveness of navigation strategies is moderated by the game's timing conditions. Specifically, in the Least-clicks games lacking a time restriction, both hub-driven and proximity-driven approaches can significantly improve performance. Conversely, in timed Speed-race games, while the hub-driven strategy remains beneficial, the proximity-driven strategy has the opposite effect. This difference may stem from the fact that it takes more time to identify pages closely related to the target, as opposed to directly jumping to a highly connected Wikipedia page. Given the negative correlation between the hub-driven and proximity-driven scores, we calculated the variance inflation factor (VIF) for each predictor in both models. Our analysis indicates that multicollinearity is not a significant concern with the maximum VIF value being 3.03.

%tab:performance
\begin{table}[h!] 
\centering 
\begin{tabular}{@{\extracolsep{5pt}}lcc} 
\\[-1.8ex]\hline 
\hline \\[-1.8ex] 
 & \multicolumn{2}{c}{\textit{Dependent variable:}} \\ 
\cline{2-3} 
\\[-1.8ex] & Seconds saved & Steps saved \\ 
 & Speed-race games & Least-click games \\ 
\hline \\[-1.8ex] 
 Steps & $-$7.066$^{***}$ (0.345) &  \\ 
  Seconds &  & $-$0.002$^{***}$ (0.0001) \\ 
  Hub-driven score & 3.248$^{**}$ (1.051) & 0.405$^{***}$ (0.041) \\ 
  Proximity-driven score & $-$3.259$^{**}$ (1.103) & 0.200$^{***}$ (0.039) \\ 
  Source page knowledge & 3.268$^{**}$ (1.116) & 0.030 (0.037) \\ 
  Target page knowledge & 1.436 (1.032) & 0.034 (0.036) \\ 
  Game Round & 1.027$^{**}$ (0.321) & $-$0.012 (0.011) \\ 
  Constant & 107.372$^{***}$ (4.010) & 3.209$^{***}$ (0.110) \\ 
 \hline \\[-1.8ex] 
Observations & 1,174 & 1,707 \\ 
R$^{2}$ & 0.367 & 0.203 \\ 
Adjusted R$^{2}$ & 0.359 & 0.196 \\ 
Residual Std. Error & 27.904 (df = 1159) & 1.182 (df = 1692) \\ 
F Statistic & 47.984$^{***}$ (df = 14; 1159) & 30.765$^{***}$ (df = 14; 1692) \\ 
\hline 
\hline \\[-1.8ex] 
\textit{Note:}  & \multicolumn{2}{r}{$^{*}$p$<$0.05; $^{**}$p$<$0.01; $^{***}$p$<$0.001} \\ 
\end{tabular} 
\caption[Regression results for the fitness of the hub-driven and proximity-driven navigation strategies, measured as the seconds saved in the Speed-race games or steps saved in the Least-clicks games. ]{Regression results for the fitness of the hub-driven and proximity-driven navigation strategies, measured as the seconds saved in the Speed-race games or steps saved in the Least-clicks games. }
\label{tab:performance} 
\end{table}  

\subsection{Individual differences}
To understand whether participants exhibit a persistent tendency towards a hub-driven or proximity-driven strategy across the nine games, we fit a fixed effects model for the hub-driven score \( H \) and the proximity-driven score \( C \) of the navigation paths, with participants' id ($Participant$) as the fixed effect. To account for other factors that might affect these scores, we included the covariates specified in Section \ref{subsec:HC}, along with two additional variables: a binary variable \( Won \), indicating whether the game was won, and \( Type \), specifying whether the game was Speed-race or Least-clicks. Lastly, we included an interaction term between $Game$ and $Won$ to account for potential correlations between these variables. To determine whether individual tendencies across the nine games are statistically significant, we also fitted a linear regression model using the same set of predictors, excluding the individual effect (\( Participant \)).  

Table \ref{tab:individual} presents the regression results of the linear regression model (1) and (3) and the fixed effects model (2) and (4) predicting the hub-driven score and proximity-driven score, respectively. As shown in the table, for both scores, the fixed effects model fits the data better than the linear model without individual effects, resulting in a $12.4\%$ and $11.4\%$ increase in the adjusted $R^2$ values for $H$ and $C$, respectively. An F-test was conducted to assess whether the inclusion of fixed effects significantly improved the model. The results indicate that the fixed effects were statistically significant, with $F(23, 5787) = 77.74, p < 0.001$ for the fixed effects model (2) predicting $H$ and $F(23, 5787) = 79.29, p < 0.001$ for the fixed effects model (4) predicting $C$, suggesting that accounting for individual effects significantly improves the model fit.

The results highlight individual differences at multiple levels. At the game level, participants who reported greater knowledge of the target article were more likely to adopt a proximity-driven strategy. In contrast, the hub-driven tendency was significant only in Speed-race games, which involve a time constraint, and not in Least-clicks games. Additionally, as participants played more rounds of the game, they were increasingly likely to adopt the hub-driven strategy, while no such learning effect was observed for the proximity-driven strategy.

At the individual level, a further analysis of individual differences suggests that personal characteristics influence navigation tendencies as well. Table \ref{tab:tendency} presents the results of a linear regression model predicting individual tendencies to use the hub-driven and proximity-driven strategies (estimated in the fixed effects model above) using participant demographic data. The results indicate that older participants tend to adopt a proximity-driven strategy, while younger participants prefer a hub-driven strategy. Additionally, white participants and female participants show a stronger inclination towards the proximity-driven strategy but do not exhibit a significant tendency—either positive or negative—towards the hub-driven strategy. In contrast, language proficiency and political stance do not appear to have a significant impact on navigation strategy. Note that the model includes the total number of games won and the number of Speed-race games played by each participant as covariates to control for their potential correlation with navigation strategy tendencies.

\subsection{Interplay of geography and occupation}
\label{subsec:interplay}

Previous research suggests that individuals primarily rely on geographic and occupational information when navigating social networks~\cite{killworth1978reversal, adamic2005search}. In this study, we examine whether similar strategies apply to navigation within knowledge networks, specifically when the target is a Wikipedia article about a person. To identify navigation paths that leverage geographic information, we mapped the visited Wikipedia articles to their corresponding Wikidata entries. Wikidata is a free, collaborative, multilingual secondary knowledge base that collects structured data to support Wikipedia and other applications~\cite{Wikidata}. We classified a Wikipedia article as geography-related if its corresponding Wikidata entry contained a coordinate location specified by the P625 property. Using this approach, we successfully mapped 11,434 (99.3\%) of the 11,511 visited articles to Wikidata, identifying 1,715 as geography-related. A navigation path is classified as \textit{geographical} if it includes at least one geography-related article in its sequence of visited pages; otherwise, it is categorized as \textit{non-geographical}.

\begin{figure}
    \centering
    \includegraphics[width=0.8\linewidth]{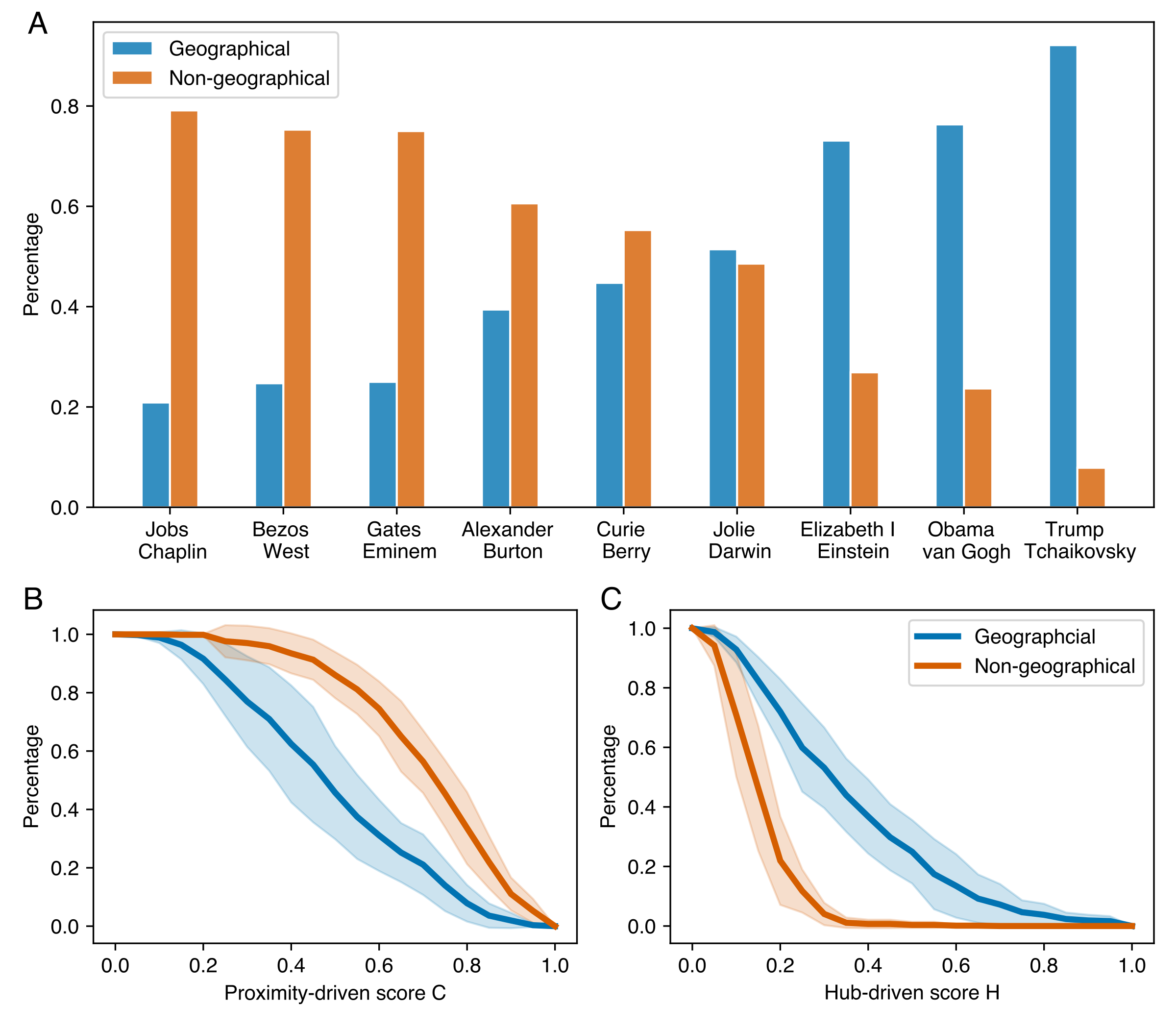}
    \caption[Figure \textbf{A} displays the percentage of geographical navigation paths across all nine games. Figure \textbf{B} illustrates the proportion of navigation paths (vertical axis) with a proximity-driven score exceeding a given threshold $C$ (horizontal axis), where the solid line represents the average across the nine games, and the error bars denote the standard deviations. The analysis is conducted separately for geographical non-geographical paths. Figure \textbf{C} follows the same format as figure \textbf{B}, but for the hub-driven score. Note that only successful navigation paths are included; for results on failed navigation paths, refer to Fig. \ref{fig:geo_lost}. ]{Comparison of the geographical and non-geographical successful navigation paths. }
    \label{fig:geo_won}
\end{figure}

Fig. \ref{fig:geo_won}A shows the percentage of successful navigation paths that are geographical. As shown, people's tendency to use geographical information to navigate or not varies across nine games, Among the successful navigation paths, a maximum 92.1\% of participants getting to the page ``Pyotr Ilyich Tchaikovsky" from ``Donald Trump" using geography-related pages, and a minimum of 20.9\% from ``Steve Jobs" to ``Charlie Chaplin". Among the paths that failed, we see similar patterns (Fig. \ref{fig:geo_lost}): the ratio of geographical paths ranges from 59.3\% to 26.2\% across nine games. We also observed a strong correlation between the ratio of geographical paths across the nine games among the successful and failed navigation paths (Pearson r=0.92).

To further understand the distinction between geographical and non-geographical navigation paths, we analyzed each navigation sequence and classified a visited Wikipedia article as a hub or proximity \textit{clue} if it had the highest hierarchical or proximity score among all visited articles. These clues reveal how participants leveraged Wikipedia’s hierarchical structure and semantic distances to navigate. Fig. \ref{fig:hc_top5} shows the five most frequently visited hub and proximity clues in each game, along with the cumulative percentage of participants who used them, for both geographical and non-geographical paths. In geographical paths, more than half of the navigation sequences (M = 0.669, SD = 0.144) can be characterized by just five hub clues, which are predominantly well-known countries and cities, such as the ``United States". The proximity clues in geographical paths vary: some participants' navigation rely heavily on geographical information, never reaching a proximity clue related to the target person’s occupation, while others combined geographical context with occupational information to reach the target. In non-geographical paths where no geography-related articles were visited, hub and proximity clues were primarily related to the target person’s profession. However, there were exceptions. For example, some participants reached Vincent van Gogh through ``Protestantism", leveraging the artist’s religious background, while others reached ``Kanye West" not only through music-related articles but also via ``Twitter" and ``Kim Kardashian", reflecting his presence in public discourse.

How do the geographical and occupational navigation strategies observed in social navigation relate to the hub-driven and proximity-driven strategies in knowledge network navigation? Fig. \ref{fig:geo_won}B and \ref{fig:geo_won}C show the proportion of navigation paths with a proximity-driven score exceeding C as a function of C (Fig. \ref{fig:geo_won}B) and the proportion of paths with a proximity-driven score above H as a function of H (Fig. \ref{fig:geo_won}C). These analyses include all successful navigation paths across nine games, for geographical and non-geographical paths respectively. Our results indicate that geographical paths are generally more hub-driven, whereas non-geographical paths tend to be proximity-driven. However, this distinction is not absolute, particularly for geographical paths. Some geographical navigation sequences also exhibit high proximity-driven scores, likely reflecting a mixed strategy that incorporates both geographical and occupational information. For failed navigation paths, Figure \ref{fig:geo_lost} shows that the differences between geographical and non-geographical paths in terms of hub-driven and proximity-driven tendencies are less pronounced. This is likely because a considerable number of paths failed to reach effective hub or proximity clues, resulting in both low hub-driven and low proximity-driven scores.

\section{Methods}\label{Methods}

\subsection{The experiment}
\label{subsec:experiment}
Our longitudinal study comprises two rounds of online experiments, the first conducted in January 2020 and the second in October 2023. Participants were sourced from Prolific~\cite{Prolific}, a well-regarded crowdsourcing platform for behavioral studies~\cite{douglas2023data}. The experiments were conducted on the Qualtrics~\cite{Qualtrics} platform, where we embedded Wikipedia navigation games into the Qualtrics survey using custom JavaScript, followed by a survey. We utilized the 20190820 English Wiki Dump~\cite{Wikidump} for the navigation games in both experiment rounds. This Wikipedia snapshot includes 5.9 million nodes and 133.6 million edges.

In our experiment, each participant plays nine rounds of the Wikipedia navigation game, followed by a survey. The source and target Wikipedia articles for each game are listed in Fig. \ref{fig:hc_paths}. In each round, participants can choose between two types of games: (1) the Speed-race game, where they must navigate to the target page within 150 seconds, and (2) the Least-clicks game, where they must reach the target in no more than seven clicks. Before the game starts, participants have 60 seconds to read a brief introduction to both the source and target Wikipedia articles. After this, they must decide whether to play the Speed-race or Least-clicks game. During each game, the interface displays pages visited earlier in the current game on the left margin, allowing players to backtrack to any of those pages (backclick). Following the game session, the survey sessions commence with a Big Five personality test~\cite{goldberg1992development}, assessing participants’ five personality traits: openness to experience, conscientiousness, extroversion, agreeableness, and neuroticism. Following this, we pose six categories of questions to gather information about participants’ i) employment status, ii) educational background, iii) spatial navigation habits, and their previous experience with iv) the Wikipedia navigation game, v) the Wikipedia website, and vi) computer games. We also inquire about demographic details, including age, gender, ethnicity, political stance, and language skills. An attention check question is included at the survey’s end, requiring participants to slide a bar to the left.

\subsection{Embedding of the Wikipedia articles}
\label{subsec:embedding}

To quantify the similarity between Wikipedia articles, we trained a 64-dimensional graph embedding for each Wikipedia page  \(a_{i}\)  across the English Wikipedia graph \(G\) using the DeepWalk algorithm~\cite{perozzi2014deepwalk}. The Wikipedia network was first converted into an undirected graph, where two articles were linked if a directed link existed between them in either directions in the original directed graph. We used the default parameters of the DeepWalk algorithm for training, setting the representation size to 64, the window size to 5, the walk length to 40, and the number of walks to 10. Unless otherwise specified, all embedding vectors were subsequently normalized to have unit length for further calculations.

To assess the effectiveness of our embedding, we conducted tests using the WikipediaSimilarity 353 Test~\cite{witten2008effective}, an adaptation of the earlier dataset, WordSimilarity 353 Test~\cite{finkelstein2001placing}, designed to evaluate semantic relatedness among words. Our graph embedding achieved a Spearman rank correlation score of 0.667 with the WikipediaSimilarity 353 test, demonstrating performance on par with the current best measures of semantic relatedness for Wikipedia pages~\cite{singer2013computing}.

\section{Discussion}\label{Discussion}

Our study investigated the navigation strategies employed by participants during navigation tasks on the Wikipedia network. We extended existing research on navigation in the information space by emphasizing individual differences—an aspect largely overlooked in prior studies, which primarily focused on aggregated navigation behavior. In addition, we drew a novel connection between information space navigation and social navigation by demonstrating that the choice to navigate by the occupational or geographical information of the target person in social navigation corresponds to the proximity-driven and hub-driven strategies identified in our analysis.

Using a graph embedding trained on the English Wikipedia network capturing the semantic distances among articles and a local hierarchical score measuring the articles' position in the Wikipedia knowledge hierarchy, we identified two navigation strategies that the participants adopted to win the game. In the hub-driven strategy participants tend to ``zoom out", looking for articles higher in the Wikipedia hierarchy to navigate, while in the proximity-driven strategy participants ``home in" at each step to articles that's semantically closer to the target. We found that such split of strategies previously discovered as the zoom-out and home-in phases in~\cite{west2012human} is not only present within one navigation game, but also at the individual level. Such individual variance is statistically significant and cannot be ignored. We analyzed the impact of demographic factors on individual tendencies to adopt hub-driven versus proximity-driven navigation strategies. Our findings indicate that age, gender, and ethnicity significantly influence these preferences. Our work connects the findings of social navigation to our navigation tasks on Wikipedia and shows that people’s different tendencies to use occupation or geography information of the target person to navigate can be understood as different choices between the hub-driven and proximity-driven strategies. 

In our experiment, we implemented social navigation within the information space, where participants' navigation trajectories reflect their thought processes rather than the people in their social networks. We observed that the division of occupational and geographical navigation paths discovered in previous work on social network navigation~\cite{milgram1967small, killworth1978reversal, dodds2003experimental} also exist in information space navigation, suggesting that representation of the geographical origin and occupation of people may be foundational to our cognitive map of the social world. Indeed, prior research has indicated that our hippocampus is capable of representing abstract quantities, such as a person's affiliations and power within social encounters~\cite{tavares2015map}, facilitating the search for suitable assistance in finding accommodation or employment.

Previous research on wayfinding in the information network~\cite{west2012human} has studied the interplay between the degree and proximity of the nodes on the network within a single navigation trajectory. Our findings extend this by showing that this interplay occurs not only in the navigation process of individual players but also at a macro level across different players. Previous models of human navigation behavior typically adopt an aggregated approach, treating individuals uniformly~\cite{helic2013models}. We hypothesize that incorporating individual variability in navigation strategies could provide a more accurate explanation of the empirical data on knowledge navigation.

Our study has several limitations that should be considered. While our experiment focused on renowned individuals as the source and target of the navigation tasks, these can be extended to lesser-known individuals or even non-human concepts, such as objects, events, or theoretical ideas. The extent to which navigation strategies differ in such contexts remains an open question. Furthermore, our navigation tasks are specific to the targeted navigation scenario, which occurs less frequently in real life compared to more general information searches, potentially limiting the generalizability of our findings. Other research has examined more realistic “navigation in the wild” scenarios, particularly within Wikipedia, by analyzing web server logs or clickstream data from online users~\cite{piccardi2023large, arora2022wikipedia}.

Our study advances prior research by revealing individual differences in information space navigation strategies and linking these strategies to mechanisms observed in social navigation. A logical progression would be to introduce navigation tasks where source and target pages are not limited to well-known individuals, but instead include lesser-known individuals or non-human concepts such as objects, events, or theoretical ideas. Furthermore, investigating algorithms to enhance online navigation support presents a promising research direction.

\bmhead{Abbreviations}

M, Mean; SD, Standard Deviation.

\backmatter

\section{Declarations}

\bmhead{Ethics approval and consent to participate}

All subjects gave their informed consent for inclusion before they participated in the study. The protocol of the study was approved by the Ethics Committee of Central European University (reference number: 2022-2023/1/EX). All methods of the study were carried out following the principles of the Belmont Report. 

\bmhead{Consent for publication}

All authors have read and approved the final manuscript for publication. Informed consent was obtained from all participants for the publication of their data. No identifiable personal data is included in this manuscript.

\bmhead{Availability of data and materials}

The datasets generated and/or analyzed during the current study are not publicly available due to ethical reasons but are available from the corresponding author on reasonable request.

\bmhead{Competing interests}

The authors declare that they have no competing interests.

\bmhead{Funding}

This project was supported by the Humboldt Foundation within the Research Group Linkage Program. JK and MZ were partially supported through ERC grant No. 810115-DYNASET. MZ is funded by the European Union under Horizon EU project LearnData, 101086712. JK acknowledges further support from Horizon 2020 ``INFRAIA-01-2018-2019" project ``SoBigData++", grant No. 871042.

\bmhead{Authors' contributions} 

All authors contributed to the conception and design of the research. MZ led the experiment and collected the data. All authors analyzed the data and wrote the paper.

\bmhead{Acknowledgments}

We are grateful to Markus Strohmaier for his valuable advice and Melanie Oyarzún for the helpful discussion. We would like to thank Csaba Pléh for his advice on the psychological literature and Peter Kardos for consultations on the technical details of the experiments.

\clearpage
\begin{appendices}

\section{Figures}\label{secA1}

This section contains the supplementary figures for the study. 

%fig:geo_lost
\begin{figure}[H]
    \centering
    \includegraphics[width=0.8\linewidth]{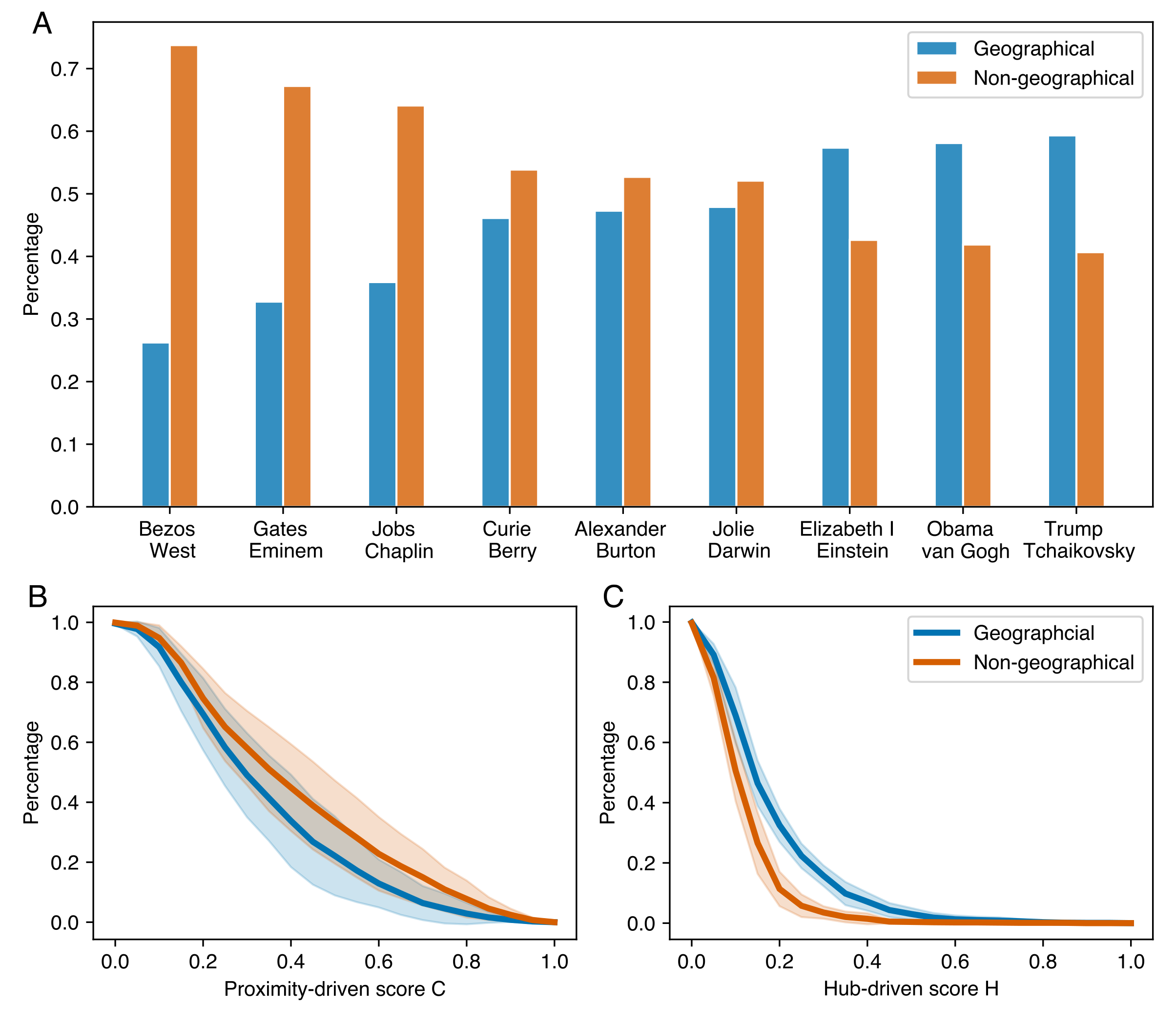}
    \caption[Figure \textbf{A} displays the percentage of geographical navigation paths across all nine games. Figure \textbf{B} illustrates the proportion of navigation paths (vertical axis) with a proximity-driven score exceeding a given threshold $C$ (horizontal axis), where the solid line represents the average across the nine games, and the error bars denote the standard deviations. The analysis is conducted separately for geographical and non-geographical paths. Figure \textbf{C} follows the same format as figure \textbf{B}, but for the hub-driven score. Note that only failed navigation paths are included; for results on successful navigation paths, refer to Fig. \ref{fig:geo_won}. ]{Comparison of the geographical and non-geographical successful navigation paths. }
    \label{fig:geo_lost}
\end{figure}

\begin{figure}[H]
    \centering
    \includegraphics[width=\linewidth]{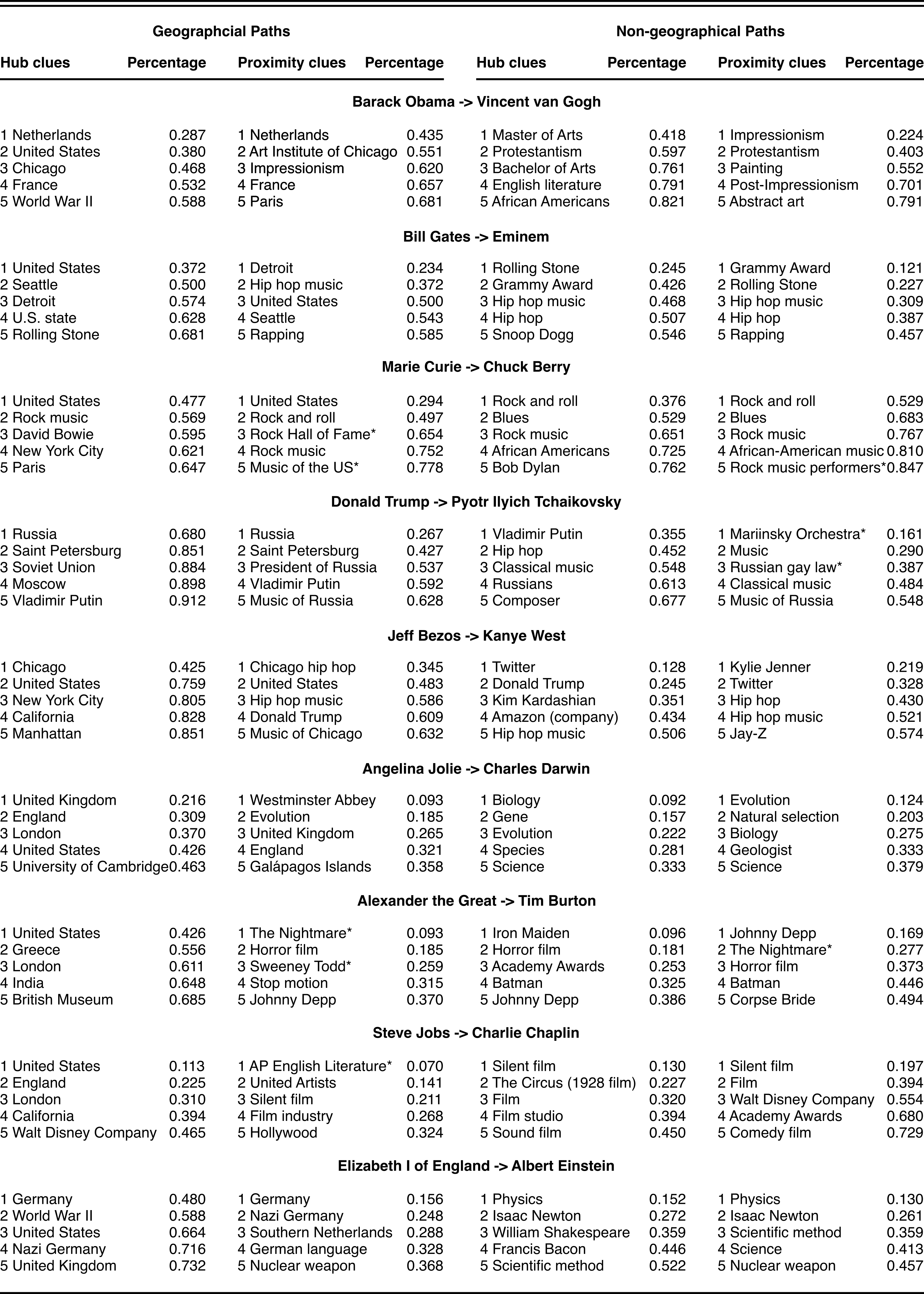}
    \caption[The figure shows the top five most visited hub and proximity clues (see definition in section \ref{subsec:interplay}) in each game, along with the cumulative percentage of participants who used them, for both geographical and non-geographical paths. Articles marked with an asterisk (*) indicate abbreviated titles of the original articles. ]{Top five most visited hub and proximity clues for all games. }
    \label{fig:hc_top5}
\end{figure}

\clearpage

\section{Tables}\label{secA2}

This section contains the supplementary tables for the study. 

%tab:individual
\begin{table}[!htbp] \centering 
\small 
\resizebox{\textwidth}{!}{%
\begin{tabular}{@{\extracolsep{5pt}}lcccc} 
\\[-1.8ex]\hline 
\hline \\[-1.8ex] 
 & \multicolumn{4}{c}{\textit{Dependent variables:}} \\ 
\cline{2-5} 
\\[-1.8ex] & \multicolumn{2}{c}{\textbf{Hub-driven Score}} & \multicolumn{2}{c}{\textbf{Proximity-driven Score}} \\ 
\\[-1.8ex] & \textit{OLS} & \textit{felm} & \textit{OLS} & \textit{felm} \\ 
\\[-1.8ex] & (1) & (2) & (3) & (4) \\ 
\hline \\[-1.8ex] 
 Won & 0.142$^{***}$ (0.011) & 0.098$^{***}$ (0.010) & 0.149$^{***}$ (0.016) & 0.109$^{***}$ (0.016) \\ 
  & & & & \\ 
 Game Round & 0.003$^{***}$ (0.001) & 0.003$^{***}$ (0.001) & $-$0.001 (0.001) & $-$0.001 (0.001) \\ 
  & & & & \\ 
 Type [Speed-race] & 0.038$^{***}$ (0.004) & 0.017$^{***}$ (0.005) & 0.003 (0.006) & $-$0.016$^{**}$ (0.007) \\ 
  & & & & \\ 
 Source page knowledge & $-$0.004$^{*}$ (0.002) & $-$0.004 (0.003) & $-$0.008$^{**}$ (0.003) & 0.002 (0.004) \\ 
  & & & & \\ 
 Target page knowledge & $-$0.003 (0.002) & $-$0.002 (0.002) & 0.010$^{***}$ (0.003) & 0.013$^{***}$ (0.003) \\ 
  & & & & \\ 
 Steps & $-$0.008$^{***}$ (0.002) & $-$0.011$^{***}$ (0.002) & $-$0.018$^{***}$ (0.003) & $-$0.007$^{**}$ (0.003) \\ 
  & & & & \\ 
 Seconds & $-$0.002 (0.002) & $-$0.003 (0.002) & 0.041$^{***}$ (0.003) & 0.021$^{***}$ (0.003) \\ 
  & & & & \\ 
 Game [2] & $-$0.008 (0.010) & $-$0.008 (0.009) & 0.035$^{**}$ (0.015) & 0.053$^{***}$ (0.014) \\ 
  & & & & \\ 
 Game [3] & 0.011 (0.010) & 0.014 (0.010) & $-$0.064$^{***}$ (0.015) & $-$0.038$^{***}$ (0.015) \\ 
  & & & & \\ 
 Game [4] & 0.036$^{***}$ (0.010) & 0.046$^{***}$ (0.010) & 0.046$^{***}$ (0.015) & 0.059$^{***}$ (0.014) \\ 
  & & & & \\ 
 Game [5] & $-$0.025$^{**}$ (0.010) & $-$0.028$^{***}$ (0.009) & 0.033$^{**}$ (0.015) & 0.049$^{***}$ (0.014) \\ 
  & & & & \\ 
 Game [6] & 0.012 (0.010) & 0.010 (0.009) & $-$0.011 (0.014) & 0.003 (0.014) \\ 
  & & & & \\ 
 Game [7] & $-$0.016$^{*}$ (0.009) & $-$0.026$^{***}$ (0.009) & 0.050$^{***}$ (0.014) & 0.055$^{***}$ (0.013) \\ 
  & & & & \\ 
 Game [8] & 0.026$^{***}$ (0.010) & 0.025$^{***}$ (0.009) & 0.138$^{***}$ (0.015) & 0.150$^{***}$ (0.014) \\ 
  & & & & \\ 
 Game [9] & 0.031$^{***}$ (0.010) & 0.032$^{***}$ (0.010) & $-$0.104$^{***}$ (0.015) & $-$0.082$^{***}$ (0.015) \\ 
  & & & & \\ 
 Won:Game [2] & $-$0.064$^{***}$ (0.015) & $-$0.053$^{***}$ (0.014) & 0.096$^{***}$ (0.022) & 0.073$^{***}$ (0.021) \\ 
  & & & & \\ 
 Won:Game [3] & $-$0.015 (0.015) & $-$0.009 (0.014) & 0.149$^{***}$ (0.022) & 0.127$^{***}$ (0.021) \\ 
  & & & & \\ 
 Won:Game [4] & 0.062$^{***}$ (0.015) & 0.055$^{***}$ (0.014) & 0.036 (0.022) & 0.024 (0.021) \\ 
  & & & & \\ 
 Won:Game [5] & $-$0.094$^{***}$ (0.015) & $-$0.082$^{***}$ (0.014) & 0.124$^{***}$ (0.022) & 0.098$^{***}$ (0.021) \\ 
  & & & & \\ 
 Won:Game [6] & $-$0.061$^{***}$ (0.015) & $-$0.052$^{***}$ (0.014) & 0.079$^{***}$ (0.022) & 0.063$^{***}$ (0.021) \\ 
  & & & & \\ 
 Won:Game [7] & $-$0.095$^{***}$ (0.017) & $-$0.073$^{***}$ (0.016) & 0.110$^{***}$ (0.025) & 0.110$^{***}$ (0.024) \\ 
  & & & & \\ 
 Won:Game [8] & $-$0.074$^{***}$ (0.015) & $-$0.064$^{***}$ (0.014) & 0.049$^{**}$ (0.022) & 0.038$^{*}$ (0.021) \\ 
  & & & & \\ 
 Won:Game [9] & 0.075$^{***}$ (0.015) & 0.082$^{***}$ (0.014) & $-$0.020 (0.022) & $-$0.033 (0.021) \\ 
  & & & & \\ 
 Constant & 0.109$^{***}$ (0.008) &  & 0.373$^{***}$ (0.011) &  \\ 
  & & & & \\ 
\hline \\[-1.8ex] 
Observations & 6,551 & 6,551 & 6,551 & 6,551 \\ 
R$^{2}$ & 0.268 & 0.461 & 0.309 & 0.489 \\ 
Adjusted R$^{2}$ & 0.266 & 0.390 & 0.307 & 0.421 \\ 
Residual Std. Error & 0.140 (df = 6527) & 0.128 (df = 5787) & 0.208 (df = 6527) & 0.190 (df = 5787) \\ 
F Statistic (full) & 104.041$^{***}$ (df = 23; 6527) & 6.496$^{***}$ (df = 763; 5787) & 127.093$^{***}$ (df = 23; 6527) & 7.253$^{***}$ (df = 763; 5787) \\ 
F Statistic (proj) &  & 77.74$^{***}$ (df = 23; 5787) &  & 79.29$^{***}$ (df = 23; 5787) \\ 
\hline 
\hline \\[-1.8ex] 
\textit{Note:}  & \multicolumn{4}{r}{$^{*}$p$<$0.1; $^{**}$p$<$0.05; $^{***}$p$<$0.01} \\ 
\end{tabular}%
}
\caption[Regression results for the linear and fixed effects models to predict the hub-driven and proximity-driven score for each navigation paths. ]{Regression results for the linear and fixed effects models to predict the hub-driven and proximity-driven score for each navigation paths. } 
\label{tab:individual} 
\end{table} 

%tab:tendency
\begin{table}[!htbp] \centering 
\begin{tabular}{@{\extracolsep{5pt}}lcc} 
\\[-1.8ex]\hline 
\hline \\[-1.8ex] 
 & \multicolumn{2}{c}{\textit{Dependent variable:}} \\ 
\cline{2-3} 
\\[-1.8ex] & Hub-driven Tendency & Proximity-driven Tendency \\ 
\\[-1.8ex] & (1) & (2)\\ 
\hline \\[-1.8ex] 
 Gender [Female] & $-$0.005 & 0.040$^{***}$ \\ 
  & (0.005) & (0.008) \\ 
  & & \\ 
 Age & $-$0.008$^{***}$ & 0.013$^{***}$ \\ 
  & (0.003) & (0.004) \\ 
  & & \\ 
 Ethnicity [White] & $-$0.0003 & 0.038$^{***}$ \\ 
  & (0.006) & (0.008) \\ 
  & & \\ 
 Native in a foreign language & $-$0.0001 & 0.019$^{*}$ \\ 
  & (0.008) & (0.011) \\ 
  & & \\ 
 Speaks a foreign language & 0.006 & $-$0.005 \\ 
  & (0.006) & (0.008) \\ 
  & & \\ 
 Political stance [Liberal] & $-$0.004 & 0.010 \\ 
  & (0.008) & (0.012) \\ 
  & & \\ 
 Political stance [Moderate] & $-$0.003 & $-$0.005 \\ 
  & (0.009) & (0.013) \\ 
  & & \\ 
 Number of games won & 0.025$^{***}$ & 0.047$^{***}$ \\ 
  & (0.003) & (0.004) \\ 
  & & \\ 
 Number of Speed-race game played & $-$0.011$^{***}$ & $-$0.001 \\ 
  & (0.003) & (0.004) \\ 
  & & \\ 
 Constant & 0.130$^{***}$ & 0.388$^{***}$ \\ 
  & (0.009) & (0.013) \\ 
  & & \\ 
\hline \\[-1.8ex] 
Observations & 698 & 698 \\ 
R$^{2}$ & 0.177 & 0.239 \\ 
Adjusted R$^{2}$ & 0.166 & 0.230 \\ 
Residual Std. Error (df = 688) & 0.068 & 0.100 \\ 
F Statistic (df = 9; 688) & 16.403$^{***}$ & 24.073$^{***}$ \\ 
\hline 
\hline \\[-1.8ex] 
\textit{Note:}  & \multicolumn{2}{r}{$^{*}$p$<$0.1; $^{**}$p$<$0.05; $^{***}$p$<$0.01} \\ 
\end{tabular} 
\caption[Regression results of linear models predicting each participant's navigation strategy tendency based on their demographic characteristics.]{Regression results of linear models predicting each participant's navigation strategy tendency based on their demographic characteristics. } 
\label{tab:tendency} 
\end{table}

\end{appendices}

\clearpage
\bibliography{sn-bibliography}

\clearpage
\listoffigures

\clearpage
\listoftables

\end{document}